\documentclass[twocolumn, showpacs, superscriptaddress, amsmath,amsfonts,prb]{revtex4}
\usepackage{graphicx}
\begin{document}

\title{Plasmonic mediated nucleation of resonant nano-cavities in metallic layers}

\author{V. G. Karpov}\email{victor.karpov@utoledo.edu}\affiliation{Department of Physics and Astronomy, University of Toledo, Toledo, OH 43606, USA}
\author{M. Nardone} \affiliation{Department of Environment and Sustainability, Bowling Green State University, Bowling Green, OH 43403, USA}
\author{A. V. Subashiev}\affiliation{Department of  Electrical and Computer Engineering, SUNY Stony Brook, Stony Brook, NY 11794 USA}
\date{\today}

\begin{abstract}
We predict plasmonic mediated nucleation of pancake shaped resonant nano-cavities in metallic layers that are penetrable to laser irradiation. The underlying physics is that the cavity provides a narrow plasmonic resonance that maximizes its polarizability in an external field.  The resonance yields a significant energy gain making the formation of such cavities highly favorable.  Possible implications include nano-optics and generation of the dielectric bits in conductive films that underlie the existing optical recording phase change technology.
\end{abstract}
\maketitle

Non-photochemical acceleration of nucleation under laser irradiation or static electric fields has been observed in a number of systems. \cite{garetz1996, nucrateE, liubin1997, nardone2012} The effect was attributed to electric field induced polarization and lowering of the nucleation barrier.  For ac fields, the  frequency dependence of the material polarization, especially near plasmonic resonances, has not yet been considered and may completely modify that effect; it is of particular importance for metallic phases.

In this letter we predict a phenomenon of plasmonic mediated nucleation of nano-cavities in metallic layers that are penetrable to laser fields. The underlying idea is that the nucleated cavity provides a narrow plasmonic resonance that maximizes its polarizability in an external field.  The ability to adjust resonance oscillations frequency and phase  makes such cavities highly energetically favorable.

Our consideration is based on the classical nucleation theory \cite{landau1980,kaschiev2000}, which accounts for the bulk $\mu V$ and surface $\sigma A$ contributions to the free energy.  With the addition of the term $F_E$ to describe the electric polarization gain, the free energy is $F=F_E+\mu V+\sigma A$, where $\mu$ is the difference in chemical potential (per volume) due to cavity nucleation, $\sigma$ is the surface tension, and $V$ and $A$ are the cavity volume and surface area, respectively.  Our analysis below starts with the case of $\mu <0$, corresponding to a metastable metallic system wherein nucleation is naturally expected.  We then consider the case of a stable metal layer, $\mu >0$, where cavities are energetically unfavorable in zero field.

For a static field, the polarization induced energy gain of a particle that nucleates in a dielectric material, with permittivity $\epsilon$, can be represented as, \cite{kaschiev2000}
\begin{equation}\label{eq:FE}
F_E=-\epsilon\alpha E^2,
\end{equation}
where  $\alpha$ is the particle polarizability, and $E$ is the field strength. A subtle point here is that $\epsilon$ makes Eq. (\ref{eq:FE}) different from the energy of a dipole in an external field; $\epsilon$ reflects the contributions from all charges in the system, including those responsible for the field.  That factor was confirmed by several authors. \cite{kaschiev2000,warshavsky1999,isard1977}

Eq. (\ref{eq:FE}) was originally obtained by integrating, over the entire space, the energy density difference caused by introduction of the particle. It must be modified for the case of dispersive media (metals), in which
\begin{equation}\label{eq:perm}
\epsilon(\omega)=1-\frac{\omega _p^2}{\omega ^2}+ i\frac{\omega_p^2}{\omega^3\tau}.
\end{equation}
where $\omega$ is the field frequency, $\tau$ is the relaxation time, and $\omega _p=\sqrt{4\pi Ne^2/m}$ is the plasma frequency with $N$ being the electron concentration, $m$ the electron mass, and $e$ the electron charge. We assume, as usual, $\tau ^{-1}\ll\omega\ll\omega _p.$ The field energy density in strongly dispersive media is given by the Brillouin formula $(\partial (\omega\epsilon )/\partial \omega )(\overline{E^2}/8\pi )$ where the overline implies a time average [see e.g. Eq. (80.12) in Ref. \onlinecite{landau1984}]. With the modification $\epsilon\rightarrow \partial (\omega\epsilon )/\partial \omega$, the derivation steps \cite{kaschiev2000,warshavsky1999,isard1977} leading to Eq. (\ref{eq:FE}) yield,
\begin{equation}\label{eq:electro}
F_E=-\frac{E^2}{2}\Re\left[\frac{\alpha\partial (\epsilon\omega )}{\partial\omega}\right],
\end{equation}
where $E$ is the field amplitude, $\Re$ represents the real part, and we have employed standard time averaging.\cite{landau1984}  Since $\omega$ exceeds reciprocal nucleation times, the above time average represents an adiabatic contribution to the energy of the atomic subsystem.

We consider spheroidal particles, for which \cite{bohren1983}
\begin{equation}\label{eq:alpha}
\alpha = \frac{V}{4\pi}\frac{\epsilon_p -\epsilon(\omega)}{\epsilon(\omega)+n(\epsilon_p - \epsilon(\omega))}.
\end{equation}
Here, $\epsilon_p$ is the dielectric permittivity of the particle (cavity), and $n$ is the depolarizing factor.

We start with the standard isothermal settings of a metastable system ($\mu <0$) wherein cavities correspond to the thermodynamically stable phase, for example, in a liquid metal or in a metal supersaturated with vacancies and/or defects. The laser beam is normally incident on a metal and nucleation of small embryos takes place in its skin depth layer. In this part of our analysis, we assume the particle spherical shape typical of the classical nucleation theory.

Close to the resonance, $|\omega -\omega _r|\ll \omega _r$, using $n=1/3$ for a sphere  gives
\begin{equation}\label{eq:real}
\Re\left[\frac{\alpha\partial (\epsilon\omega )}{\partial\omega}\right]=R^3 c \frac{\left(\omega^2-\omega_r^2\right)-a \omega_r^4/(2 c)}{\left(\omega^2-\omega_r^2\right)^2+a\omega_r^4}
\end{equation}
where $R$ is the radius of the sphere,
\begin{equation}\label{eq:omegar}
a=\left(\frac{1}{\omega\tau}\right)^2, \quad c=\frac{\left(\epsilon_p-1\right)\omega^2+\omega_p^2}{\epsilon_p+2},\quad \omega_r=\sqrt{\frac{2}{2+\epsilon_p}}\omega_p.
\end{equation}
Assuming as usual $\omega\tau\gg 1$, we observe the resonance at $\omega^2\approx \omega_r^2(1-\sqrt{a})$ with a sharp minimum given by,
\begin{equation}\label{eq:alphaminsphere}
\Re\left[\frac{\alpha\partial (\epsilon\omega )}{\partial\omega}\right]_{min}\approx -\frac{R^3\epsilon_p\omega_p\tau}{2\sqrt{2+\epsilon_p}}.
\end{equation}
This is by the quality factor $Q=\omega\tau\gg 1$ greater than the static polarizability ($\sim R^3$) of a metallic sphere.

The system free energy can be written as,
\begin{equation}\label{eq:free}
F=(E^2/2)\left[\Re (\epsilon\alpha )\right]_{min}+V\mu +A\sigma,
\end{equation}
with $V=4\pi R^3/3$ and $A=4\pi R^2$. It reduces to its standard form of the classical nucleation theory with the renormalization,
\begin{equation}\label{eq:renormmu}
\mu \rightarrow \mu _E=\mu +\delta\mu, \quad \delta\mu =-|\mu |\frac{E^2R_0^3}{W_0}\omega_p\tau\frac{\pi\epsilon_p}{6\sqrt{2+\epsilon_p}}
\end{equation}
applicable both for $\mu <0$ and $\mu >0$. Here we have introduced, for convenience, the radius and barrier of the classical nucleation theory,
$$R_0=\frac{2\sigma}{|\mu |},\quad \mathrm{and}\quad W_0=\frac{16\pi \sigma ^3}{3\mu ^2}.$$
Their ballpark values are $W_0\sim 1$ eV and $R_0\sim 1$ nm for the typical cases of nucleation in solids. \cite{kaschiev2000}

\begin{figure}[htb]
\includegraphics[width=0.35\textwidth]{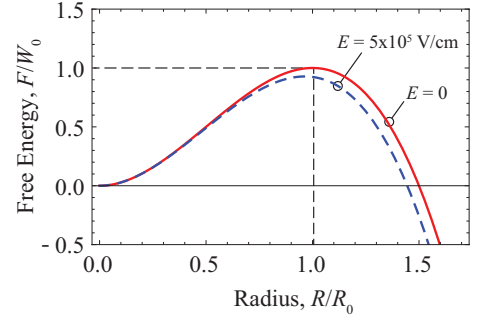}
\caption{Free energy of a spherical cavity in a metal.  A slight lowering of the barrier requires $E=4\times 10^5$ V/cm at the plasmon frequency of $\omega_r=\sqrt{2/3} \omega_p$. Parameters values are typical for void nucleation in metals:\cite{was2007} $\sigma=1$ J/m$^2$, $R_0=0.5$ nm, $W_0=1$ eV, $\omega_p=10^{15}$ rad/s, and $\tau=10^{-12}$ s, with $\epsilon_p=1$. \label{Fig:sphereplot}}
\end{figure}

We observe from Eq. (\ref{eq:renormmu}) that ac fields lower the barrier and radius of spherical void nucleation; the smallness of the effect is described by the dimensionless parameter $\xi =E^2R_0^3/W_0$ (see Fig. \ref{Fig:sphereplot}). A similar conclusion for dc fields has been long known. \cite{kaschiev2000} A feature added here is the resonant nature of the ac field effect that is significantly amplified by the $Q$-factor.

\begin{figure}[b]
\includegraphics[width=0.35\textwidth]{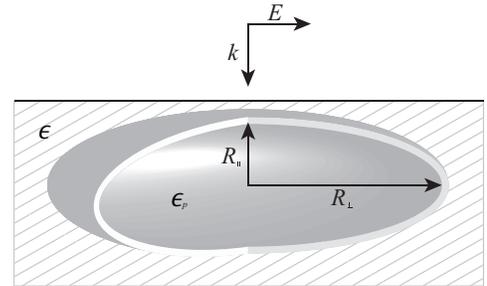}
\caption{Oblate spheroidal cavity of permittivity $\epsilon_p$, semi-minor axis $R_{\parallel}$, and semi-major axis $R_{\perp}$ embedded in a thin metal film of permittivity $\epsilon$. \label{Fig:oblate3d}}
\end{figure}

Consider next the case of ac fields strong enough to distort the spherical geometry. It is known\cite{rayleigh} that oblate spheroids remain stable with respect to elastic and electrical perturbations.  Hence, we assume an oblate spheroid with the semi-major axis $R_{\perp}$ and semi-minor axis $R_{\parallel}$ directed perpendicular to and along the direction of beam propagation, respectively (see Fig. \ref{Fig:oblate3d}).  The depolarizing factor is given by, \cite{bohren1983}
\begin{equation}\label{eq:n}
n=\frac{1+\eta ^2}{\eta ^3}(\eta -\arctan \eta )\approx 1-\frac{\pi}{2\eta}\equiv 1-\delta n,
\end{equation}
with eccentricity, $\eta=R_{\perp}/R_{\parallel}\gg 1.$ Inserting Eqs. (\ref{eq:n}) and (\ref{eq:perm}) into Eq. (\ref{eq:alpha}) and assuming $\omega\ll \omega _p$ yields,
\begin{equation}\label{eq:realalpha}
\Re\left[\frac{\alpha\partial (\epsilon\omega )}{\partial\omega}\right]=-\frac{V\epsilon _p}{4\pi \delta n_{\omega}}\frac{(\delta n_{\omega}-\delta n)-a\delta n}{(\delta n_{\omega}-\delta n)^2+a\delta n^2}
\end{equation}
where,
$$\delta n_{\omega}=\frac{\epsilon_p\omega^2}{\omega^2\left(\epsilon_p-1\right)+\omega_p^2}\approx\epsilon_p\frac{\omega ^2}{\omega _p^2}\ll 1.$$

The polarizability exhibits resonance behavior near $\delta n\approx \delta n_{\omega}$ with a sharp minimum when,
\begin{equation}\label{eq:deltan}
\delta n \approx \delta n_{\omega}(1-\sqrt{a}).
\end{equation}
The corresponding value is,
\begin{equation}\label{eq:alphamin}
\Re\left[\frac{\alpha\partial (\epsilon\omega )}{\partial\omega}\right]_{min}\approx -\frac{V}{8\pi}\frac{\omega\tau}{\epsilon_p}\left(\frac{\omega_p}{\omega}\right)^4.
\end{equation}
Eq. (\ref{eq:deltan}) shows that the aspect ratio of the nucleated void is governed by the frequency,
\begin{equation}\label{eq:plasmon}\frac{R_{\parallel}}{R_{\perp}}\approx\epsilon_p\frac{\omega^2}{\omega_p^2}.\end{equation}
$R_{\perp}$ remains to be determined by minimizing the free energy.

The resonant frequency $\omega$ expressed in Eq. (\ref{eq:plasmon}) through the aspect ratio is known as the plasmonic resonance frequency which describes collective oscillations of quasi-free electrons; it has been experimentally observed in light scattering by nanoparticles.\cite{maier2007}  As a simple qualitative argument, consider an oblate spheroidal cavity with $\epsilon _p=1$.  Shifting the positive and negative components in its surrounding plasma over small distance, $x\ll R_{\perp}$ along $R_{\perp}$, deposits charges $q\sim R_{\parallel}R_{\perp}xNe$ on the two halves of the spheroid.  Each of them exert forces $\sim qe/R_{\perp}^2$ on individual electrons on the opposite side. Interpreting the latter as the restoring forces $m\omega ^2x$ yields the resonant frequency $\omega\sim \omega _p\sqrt{R_{\parallel}/R_{\perp}}$, consistent with Eq. (\ref{eq:deltan}).

It is interesting to note that $|\left[\Re (\epsilon\alpha )\right]_{min}|$ in Eq. (\ref{eq:alphamin}) is by the $Q$-factor greater than the static polarizability of a metallic {\it prolate} spheroid of the reciprocal aspect ratio, \cite{landau1984,nardone2012} $\alpha \approx (V/8\pi)(R_{\parallel}/R_{\perp})^2$ where $R_{\parallel}/R_{\perp}\gg 1$.
In the mean time, slightly modifying the above analysis shows that a prolate spheroidal cavity in a metal does not possess any strong polarizability. This can be attributed to strong metal screening of the polarization charges induced at the spheroid  poles. The case of plasmonic driven metal prolate spheroid nucleation in dielectric media will be presented elsewhere.

\begin{figure}[htb]
\includegraphics[width=0.40\textwidth]{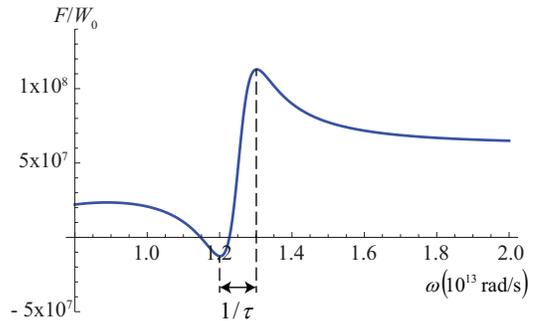}
\caption{Normalized free energy of an oblate spheroidal cavity vs. field frequency $f=\omega /2\pi$, with $E=3\times 10^3$ V/cm. The sharp resonance of width $1/\tau$ determines the aspect ratio $R_{\parallel}/R_{\perp}\approx (\omega/\omega_p)^2\approx 10^{(-4)}$.  Numerical values are the same as those used in Fig. \ref{Fig:sphereplot}. \label{Fig:freeoblateplot}}
\end{figure}

As illustrated in Fig. \ref{Fig:freeoblateplot}, the minimum in $\Re\left[\alpha\partial (\epsilon\omega )/\partial\omega\right]$ of Eq. (\ref{eq:alphamin}) is so sharp that all other terms containing $\delta n$ in the free energy of Eq. (\ref{eq:free}), with $\mu >0$, can be evaluated at $\delta n=\delta n_{\omega}$. The volume and area are then,
$$V=4\pi R_{\perp}^3/3\eta\approx 8R_{\perp}^3\delta n/3,\quad\textrm{and}\quad A=2\pi R_{\perp}^2.$$
Normalizing the free energy with respect to the classical barrier, it takes the form,
\begin{equation}\label{eq:freenorm}
\frac{F}{W_0}=\frac{4}{\pi}\frac{R_{\perp}^3}{R_0^3}\frac{\epsilon_p\omega^2}{\omega_p^2}\left[-\frac{E^2 R_0^3}{24 W_0}\left(\frac{\omega _p}{\omega}\right)^4\frac{\omega\tau}{\epsilon_p}+1\right]+\frac{3}{2}\frac{R_{\perp}^2}{R_0^2},
\end{equation}
Comparing this with Eq. (\ref{eq:freenorm}) shows that the field effect is much stronger for highly anisotropic oblate spheroids than for spheres when $\omega \ll \omega _p$.

The instability takes place when the bulk chemical contribution [second term in Eq. (\ref{eq:renormmu})] is smaller that the field term, i.e., when
\begin{equation}\label{eq:fieldcrit}E> E_c\frac{1}{\sqrt{\omega\tau}}\frac{\omega^2}{\omega_p^2},\quad E_c\equiv 2\sqrt{\frac{6 W_0\epsilon_p}{R_0^3}}.\end{equation}
Assuming the above ballpark parameter values and $\epsilon _p=1$ yields the characteristic field $E_c\sim 10^8$ V/cm. However, the other multipliers can easily make the right hand side in the inequality of Eq. (\ref{eq:fieldcrit}) much lower, taking it down to say, 1 kV/cm, which corresponds to low power lasers $P\sim 10$ mW/$\mu$m$^2$ (such as those used with, e.g, DVD burners).

Minimizing the free energy with respect to $R_{\perp}$ yields the nucleation radius and barrier,
\begin{equation}\label{eq:perrad}
R_{\perp ,0}=R_0\frac{\pi \omega _p^2}{2\omega ^2\epsilon_p}\left[\frac{E^2}{E_c^2}\left(\frac{\omega _p}{\omega}\right)^4\omega\tau-1\right]^{-1}
\end{equation}
and
\begin{equation}\label{nucbarper}
W=\frac{W_0}{2}\left(\frac{R_{\perp ,0}}{R_0}\right)^2.
\end{equation}

As a prototype system to which the above theory applies we consider again a metal close to a phase transition at temperature $T_m$, yet stable.  This enables one to estimate its chemical potential as $\mu =\mu _0(1-T/T_m)$. Correspondingly, the classical nucleation radius and barrier become,
\begin{equation}\label{eq:param}
R_0=R_{00}(1-T/T_m)^{-1},\quad W_0=W_{00}(1-T/T_m)^{-2},
\end{equation}
and $E_c=E_{c0}(1-T/T_m)^{1/2}$, where $R_{00}$, $W_{00}$ and $E_{c0}$ are estimated as the atomic scale values. The convenience of this model is that it allows macroscopically large $R_0$, consistent with the classical nucleation theory.

Using the latter relations and assuming that the first term in parenthesis of Eq. (\ref{eq:perrad}) dominates, one gets,
\begin{equation}
W=W_{00}\frac{\pi ^2}{8}\frac{1}{(\omega \tau )^2}\left(\frac{\omega}{\omega _p}\right)^4\left(\frac{E_{c0}}{E}\right)^4.
\end{equation}
Note that the nucleation barrier turns out to be temperature independent. This is in striking difference with the classical nucleation theory, which predicts a diverging nucleation barrier $W_0$ towards the phase transition temperature in Eq. (\ref{eq:param}).

We now briefly discuss possible implications of our theory. The above introduced system of a metal close to a phase transition would be most suitable for experimental verification.  Since the cavities lie within the skin layer depth, they are optically accessible and can be identified via the unique features of oblate spheroids in light scattering and absorption. \cite{grigorchuk2012} We note that such cavities can grow during the laser pulse duration by further absorbing the electromagnetic power to achieve sizes $R_{\perp}$ comparable to the beam diameter.

For stronger electric fields, the nucleation radii $R_{\perp}$ and $R_{\parallel}$ in Eqs. (\ref{eq:perrad}) and (\ref{eq:plasmon}) can shrink below the atomic length scale, which is beyond the range of the proposed macroscopic description.  However, the prediction of energy gain due to resonant cavity nucleation remains valid in the range of macroscopic post-nucleation dimensions, though the transitions rates to the lower energy state are difficult to estimate.

We note the observed formation of pancake shaped voids in metal films under moderate power density ($P\sim 1$ W/$\mu {\rm m}^2$) laser beams. \cite{georgiev2011} Similar nanovoids have been explained by thermal instabilities and recrystallization. \cite{nanovoids} The phenomenon of laser ablation \cite{ablation} is observed for $P\gtrsim 10$ W/$\mu {\rm m}^2$. The effects proposed in this Letter can be uniquely identified under much lower power densities $P\lesssim 1- 100$ mW/$\mu {\rm m}^2$ insufficient to cause melting and recrystallization instabilities.

The possibility of creating all-metal nano-cavities with well controlled shapes opens a venue towards high Q-factor resonators, which would be important in the field of nano-optics and capable of efficient lasing. \cite{yin2012,oulton2012,kim2012}

For the case of intense enough laser beams, light absorption and heating may become important concomitant factors facilitating the nucleation. Furthermore, semiconductor films under well absorbed laser light can undergo significant degree of ionization making their local (under beam) conductivities metallic and corresponding plasma frequencies exceeding the beam frequency. Our theory predicts  dielectric void formation in such semiconductor materials.

Our theory can help to elucidate the physics of optical recording in DVD and related technologies where information is kept in the form of small dielectric amorphous bits embedded in a semi-metal crystalline film (polycrystalline Ge$_2$Sb$_2$Te$_5$, etc.). The mainstream understanding has been that such metastable bits appear due to quenching of laser generated melted spots. However, recent experimental work \cite{kolobov2010} has shown that the process does not evolve through the melt. Our predicted nucleation of dielectric cavities suggests that the observed dielectric bits in semi-metal films can be created under laser irradiation (and stay as metastable inclusions afterwards) simply because they are more energetically favorable. A theory of nucleation and growth of such bits can be used to optimize optical recording with respect to the material and laser beam parameters.

In conclusion, we have predicted a phenomenon of plasmonic mediated nano-cavity nucleation in metallic layers. It may have important practical implications. More work is called upon to describe the growth stage of such resonant cavities and relate them to the experimental observations.


\begin{thebibliography}{99}
\bibitem{garetz1996} B. A. Garetz, J. E. Aber, N. L. Goddard, R. G. Young, and A. S. Myerson, Phys. Rev. Lett. {\bf 77}, 3475 (1996); B. A. Garetz, J. Matic, and A. S. Myerson, Phys. Rev. Lett. {\bf 89}, 175501 (2002); M. R. Ward, S. McHugh and A. J. Alexander, Phys. Chem. Chem. Phys. {\bf 14}, 90 (2012).
\bibitem{nucrateE} R.C. deVekey and A.J. Majumdar, Nature {\bf 225}, 172 (1970);
W. Liu, K.M. Liang, Y.K. Zheng, S.R. Gu, H. Chen, J. Phys. D Appl. Phys {\bf 30}, 3366 (1997);
J. Duchene, M. Terraillon, P. Paily, and G. Adam, Appl. Phys. Lett. {\bf 19}, 115 (1971);
B.-J. Kim, Y. W. Lee, B.-G. Chae, S. J. Yun, S.-Y. Oh, and H.-T. Kim, Appl. Phys. Lett. {\bf 90}, 023515 (2007);
K. Okimura, N. Ezreena, Y. Sasakawa, and J. Sakai, Japan J. Appl. Phys. {\bf 48}, 065003 (2009).
\bibitem{liubin1997} V. Lyubin, M. Klebanov, M. Mitkova and T. Petkova, Appl. Phys. Lett. {\bf 71}, 2118 (1997). V.I. Mikla, I.P. Mikhalko, and V.V. Mikla, Materials Science and Engineering {\bf B83}, 74 (2001).
\bibitem{nardone2012}V. G. Karpov, Y. A. Kryukov, I. V. Karpov, and M. Mitra, Phys. Rev. B {\bf 78}, 052201 (2008); I. V. Karpov, M. Mitra, G. Spadini, U. Kau, Y. A. Kryukov, and V. G. Karpov, Appl. Phys. Lett. {\bf 92}, 173501 (2008); M. Nardone and V. G. Karpov, Appl. Phys. Lett. {\bf 100}, 151912 (2012).
\bibitem{landau1980}L. D. Landau and E. M. Lifshitz, {\it Statistical Physics} 3rd edn (Pergamon, Oxford, 1980).
\bibitem{kaschiev2000}D. Kaschiev, {\it Nucleation: Basic Theory with Applications} (Butterworth-Heinemann, Oxford, 2000).
\bibitem{warshavsky1999}V.B. Warshavsky, A.K. Shchekin, Colloids and Surfaces A: Physicochemical and Engineering Aspects {\bf 148}, 283 (1999).
\bibitem{isard1977}J. O. Isard, Phil. Mag. {\bf 35}, 817 (1977).
\bibitem{landau1984} L. D. Landau, I. M. Lifshitz, and L. P. Pitaevskii, \emph{Electrodynamics of Continuous Media} (Pergamon, Oxford, New York, 1984).
\bibitem{bohren1983}C. F. Bohren and D. R. Huffman, {\it Absorption and Scattering of Light by Small Particles} (Wiley, New York 1983).
\bibitem{was2007} G. S. Was, {\it Fundamentals of Radiation Materials Science} (Springer, New York, 2007).
\bibitem{rayleigh}V. V. Voronkov and R. Falster, J. Appl. Phys. {\bf 89}, 5965 (2001); D. J. Srolovitz and S. A. Safran, J. Appl. Phys. {\bf 60}, 247 (1986); S. I. Shchukin and A. I. Grigor�ev, Technical Physics {\bf 43}, 1314 (1998).
\bibitem{georgiev2011}J. P. Moening, D. G. Georgiev, and J. G. Lawrence, J. Appl. Phys. {\bf 109}, 014304 (2011).

\bibitem{maier2007}S. A. Maier, {\it Plasmonics: Fundamentals and Applications} (Springer, New York, 2007).

\bibitem{grigorchuk2012}N. I. Grigorchuk, Europhys. Lett. {\bf 97}, 45001 (2012);
P. M. Tomchuk and N. I. Grigorchuk, Phys. Rev. B {\bf 73}, 155423 (2006).

\bibitem{nanovoids}S. I. Ashitkov, N. A. Inogamov, V. V. Zhakhovskii, Yu Emirov, M. B. Agranat, I. I. Oleinik, S. I. Anisimov, V. E. Fortov, JETP Letters {\bf 95}, 176 (2012).
\bibitem{ablation}V. M. Kozhevin, D. A. Yavsin, V. M. Kouznetsov, V. M. Busov, V. M. Mikushkin, S. Yu. Nikonov, S. A. Gurevich, A. Kolobov, J. Vac. Sci. Technol. B {\bf 18}, 1402 (2000); T. Scholz, K.Dickmann, H.Uphoff, L.Lammers, Optics and Lasers in Engineering {\bf 50}, 717 (2012).
\bibitem{yin2012}Y. Yin, T. Qiu, J. Li, P. K. Chu, Nano Energy {\bf 1}, 25 (2012).
\bibitem{oulton2012}R. F. Oulton, Materials Today {\bf 15}, 26 (2012).
\bibitem{kim2012}S-H. Kim, J. Huang, A. Scherer, Optics Letters {\bf 37}, 488 (2012).
\bibitem{kolobov2010}P. Fons, H. Osawa, A. V. Kolobov, T. Fukaya, M. Suzuki, T. Uruga, N. Kawamura, H. Tanida, and J. Tominaga, Phys. Rev. B {\bf 82}, 041203(R) (2010).


\end{thebibliography}
\end{document}